# Gravitinos, the Lithium problem, and DM production: Is there a corresponding neutrino physics linkage?

A.W. Beckwith
beckwith@aibep.org, abeckwith@uh.edu

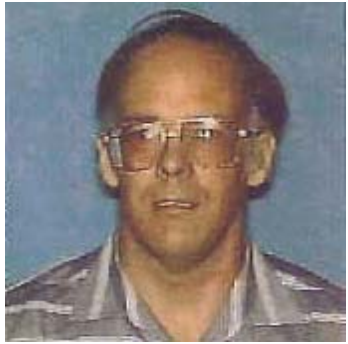

Studies are cited indicating that gravitino production acts as a natural upper bound to $Li^6$ and $Li^7$ levels, based on what happens after hadronic decay of relic 1 TeV into 100 GeV gravitinos at 1000 s. after the Big Bang. The produced gravitinos contribute a large fraction of required dark matter density. Whether or not gravitinos can be linked to neutrino production depends on which model of dark matter (DM) is assumed or used. A model presented by the author in 2008 links DM of about 100 GeV -- based on a phenomenological Lagrangian creating different Neutrino masses without SUSY -- with a dark matter candidate of about 100 GeV. This may tie in 100 GeV gravitinos with neutrino physics.

## Introduction

The author has presented (Beckwith, 2008a and 2008b) arguments relating the number of computational operations for production of entropy -- given by Seth Lloyd (2001) and modified by Beckwith (2009) -- to graviton production:

$$I = S_{total} / k_B \ln 2 = [\#operations]^{3/4} = [\rho \cdot c^5 \cdot t^4 / \hbar]^{3/4} \qquad (1)$$

Where I is total entropy divided by Boltzman's constant, $k_B$ is Boltzmann's constant, $S_{total}$ is the entropy generated by emergent space time up to a time t, and $\rho$ density is the "time component" of the usual stress-energy expression of general relativity $\rho \equiv T^{00}$. This formulation of entropy provides a way to obtain a numerical count (at or before 1000 s. after the Big Bang) of gravitons ($\Delta S \approx \Delta N_{gravitons} \approx 10^{20}$) as part of an emergent field representation of gravity/gravitational waves, a starting point for determining increasing net universe cosmological entropy. To tie in the use of Eqn. (1) with dark matter/neutrino physics, it should be noted that gravitinos (the SUSY "partner" of gravitons) are modeled by Karsten Jedamzik et al. (2008), as having a mass of 100 GeV at 1000 seconds after the Big Bang. The author's model of DM (2009) also estimated, via a non-SUSY Lagrangian argument, a mass range on the order of 100 to 400 GeV. If there is an early universe production of gravitinos as a super partner to gravitons, suppression of $Li^6$ levels is assumed to be linked to the relic production of 100 GeV gravitinos. If $Li^6$ levels are also linked to DM mass values, this may say something about relic neutrino data sets, especially if Beckwith's (2009) linkage of the Meissner and Nicholai Lagrangian for neutrino physics and DM -- using Ng's equivalence between entropy and numerical production values (2007) – is confirmed.

# Models of suppression of $Li^6$ and $Li^7$ levels due to 100 GeV gravitino generation

Jedamzik (2008, page 7) estimates that the suppression of $Li^7$ is linked to gravitinos, based on the idea that supersymmetry relates a boson to a fermion. The lack of experimental evidence of, say, a selectron (bosonic particle having all the properties of an electron except that it has zero spin) suggests that supersymmetry is broken. This selectron could then acquire large mass corrections, which would have prevented us from finding it thus far. If there is, say, a change in entropy, and the number of relic, emergent gravitational field "gravitons" from the Big Bang (the number is defined as $\Delta S \approx \Delta N_{gravitons} \approx 10^{20}$ within 1000 seconds after the Big Bang) and supersymmetry creates a similar number of super partner gravitinos, each of mass of about 100 GeV, Beckwith (2009) proposes (assuming each gravitino is paired with a relic graviton) that the number of relic neutrinos is roughly equivalent to the number of relic gravitinos. This is in line with $\Delta S \approx \Delta N_{gravitons} \approx 10^{20}$ in the first 1000 s. after the big bang. Does this lead to limits on Lithium 6 and Lithium 7 production?

## Conclusion: A certain number of gravitons/gravitinos produced leads to Lithium 6 and 7 numerical production. Does this imply relic neutrino data sets?

If a certain number of neutrinos of mass of at least 28 to 100 GeV is produced, as implied by G. Belanger (2004), the following needs to be investigated: is there roughly a one-to-one correspondence between gravitinos, neutrinos, and relic gravitons, leading to $\Delta S \approx \Delta N_{gravitons} \approx 10^{20}$ in the first 1000 s.? And if true, are there enough gravitinos and neutrinos to account for Jedamzik's (2008) data, indicating suppression of Lithium 6 and 7? The payoff of this linkage would be linking relic neutrinos and relic gravitons – leading to possibly linking the data sets of the IceCube neutrino telescope with the NIST experiment on primordial early universe gravitational waves (2009). If this linkage is confirmed, it could indicate that there are, indirectly through supersymmetry, enough neutrino-gravitino "objects" produced in the early universe to confirm the suppression of Lithium 6 and 7. It should be noted that this investigation would be more sophisticated than what McElrath (2007) suggested for limits to dark matter searches, which would require LHC style accelerator searches.  .